\documentclass[onecolumn, superscriptaddress,nofootinbib,floatfix, amsfonts, aps]{revtex4}

\usepackage{amsmath}
\usepackage{bm}
\usepackage{graphicx}
\usepackage{mathrsfs}
\usepackage{footmisc}
\usepackage{multirow}
\usepackage{graphicx}
\usepackage{booktabs, ctable}
\usepackage{xcolor, color, framed}

\begin{document}

\title{
Theoretical Review of Charmonium Production with Different $p_T$ in the Hot Medium}

\author{Baoyi Chen}
\email{baoyi.chen@tju.edu.cn} 
\affiliation{Department of Physics, Tianjin University, Tianjin 300350, China}
\affiliation{Institut f\"ur Theoretische Physik, Goethe-Universit\"at Frankfurt,
Max-von-Laue-Str. 1, D-60438 Frankfurt am Main, Germany}
\author{Carsten Greiner}
\email{carsten.greiner@th.physik.uni-frankfurt.de}
\affiliation{Institut f\"ur Theoretische Physik, Goethe-Universit\"at Frankfurt,
Max-von-Laue-Str. 1, D-60438 Frankfurt am Main, Germany}

%\date{\today}

\begin{abstract}
  Charmonia with different transverse momentum $p_T$ usually comes from different mechanisms 
in the relativistic heavy ion collisions. This work tries to review the theoretical studies on 
quarkonium evolutions in the deconfined medium produced in p-Pb and Pb-Pb collisions. The charmonia with high 
$p_T$ are mainly from the initial hadronic collisions, and therefore sensitive to the initial energy density of the 
bulk medium. For those charmonia within $0.1<p_T<5$ GeV/c at the energies of Large Hadron Collisions (LHC), 
They are mainly produced by the recombination of charm and anti-charm quarks in the medium. 
In the extremely low $p_T\sim 1/R_A$ ($R_A$ is the nuclear radius), additional contribution from the coherent 
interactions between electromagnetic fields generated by one nucleus and the target nucleus plays a 
non-negligible role in the $J/\psi$ production even in semi-central Pb-Pb collisions. 

\end{abstract}
%\pacs{25.75.Dw, 12.38.Mh, 24.85.+p}

\maketitle

\section{Introduction}
\label{intro}
Since charmonium abnormal suppression proposed as one of the signals of the deconfined medium produced 
in nucleus-nucleus (AA) collisions~\cite{Matsui:1986dk}, 
abundant theoretical studies~\cite{Grandchamp:2001pf,Andronic:2003zv,Yan:2006ve,Chen:2012gg, Blaizot:2015hya,Yao:2017fuc,Yao:2018nmy} 
about charmonium evolutions in hot 
medium have been done to reveal different aspects of heavy quark (HQ) interactions with the bulk medium. 
This paper briefly lay out theoretical models and compare them with the 
experimental data, trying to give a full picture of charmonium evolutions in the hot medium.  
Heavy quark interactions inside $c\bar c$ can be partially screened by the thermal partons 
with color charges~\cite{Burnier:2014ssa}. 
As the heavy quark interactions at large distance are screened at first, charmonium 
states ($J/\psi$, $\chi_c$, $\psi(2S)$) with different geometry sizes  
are sequentially melt as the medium temperature increases~\cite{Satz:2005hx}. 
Close to the critical 
temperature $T_c$ of the deconfined phase transition, $J/\psi$ wave function is less modified by the thermal medium 
compared with the excited states.  
The medium-induced large decay rate suppresses the production of excited states and also $J/\psi$ total 
production, as $\sim 40\%$ of the final $J/\psi$ come from the decay of excited states ($\chi_c$, $\psi(2S)$) in heavy 
ion collisions~\cite{Chen:2015iga}. 
Nuclear modification factor $R_{AA}$, defined as the ratio of charmonium yield in nucleus-nucleus (AA) collisions 
$N_{AA}^{J/\psi}$ and the production in proton-proton (pp) collisions scaled by the number of binary 
collisions $N_{coll}N_{pp}^{J/\psi}$, measures the magnitude of hot medium effects on charmonia.  
The deviation of $R_{AA}$ from unit is due to the parton inelastic scatterings and color screening effects. 

\begin{figure}
\centering
\includegraphics[width=4.0in]{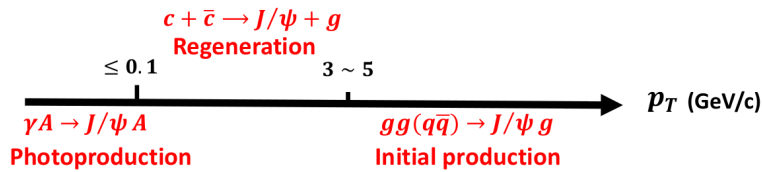}
\label{fig-schematic}
%\begin{minipage}[t]{0.43\linewidth}
%\centering
%\includegraphics[width=2.0in]{fig2}
%\label{fig2}
%\end{minipage}
\caption{Schematic figure for charmonium initial production, regeneration, coherent photoproduction dominated in 
different transverse momentum $p_T$. 
The figure is cited from~\cite{Shi:2017qep}. }
\end{figure}

With only primordial production, $J/\psi$ nuclear modification factor $R_{AA}$ 
should decrease with the colliding energies 
$\sqrt{s_{NN}}$ where hotter medium medium is expected in the larger $\sqrt{s_{NN}}$. 
However, from Relativistic Heavy Ion Collider (RHIC) to the Large Hadron Collider (LHC), 
colliding energy increases about 10 times, but the $J/\psi$ nuclear modification factor is enhanced instead of 
being suppressed~\cite{Adare:2006ns,Adam:2016rdg}. 
Besides, this enhancement is only located in the small transverse momentum region. 
This phenomina is now well clarified by the regeneration mechanism~\cite{Thews:2000rj}, 
where single charm and anti-charm quark from different pairs can combine into a new charmonium during the QGP 
expansion, due to the large binding energy of $J/\psi$ in the thermal medium. This process happens after the 
primordial production. Charmonium yield from regeneration process is proportional to the densities of charm 
and anti-charm quark $\rho_{c(\bar c)}(t,{\bf r},{\bf p}_{c(\bar c)})$ and also the cross section of the reaction 
$c+\bar c\rightarrow J/\psi +g$~\cite{Chen:2018kfo}. 
As the cross section of charm pair $\sigma_{pp}^{c\bar c}$ 
increases significantly with $\sqrt{s_{NN}}$, 
abundant charm pairs at LHC energies can significantly enhance the regeneration process and the $R_{AA}$. 
Different from color singlet bound state, single charm quarks carry color charges and are strongly coupled with 
QGP. This is supported by the observables of open charm mesons such as 
elliptic flows of D mesons and its nuclear modification factor~\cite{Adare:2006nq}. 
With strong energy loss of heavy quarks in the thermal medium~\cite{Yao:2018sgn}, 
the regenerated charmonia tend to carry small 
momentum compared with the primordially produced charmonia. Assuming charm quarks reach kinetic equilirbium 
in their momentum distribution, one can roughly obtain the 
mean transverse momentum of the regenerated charmonia $\langle p_T^\Psi\rangle \sim 2p_c\sim 4T_{QGP}$. 
At RHIC and LHC Pb-Pb 
collisions, QGP typical temperature in the most phase space (not the initial maximum temperature) is around 
$T_{QGP}\sim 200-300$ MeV, which results in $\langle p_T^\Psi\rangle \sim 1$ GeV/c for regenerated charmonium. 
It is below the mean transverse 
momentum of the primordially produced charmonia $2\sim 3$ GeV/c.   
As charm quarks are strongly coupled with the anisotropic medium, 
they also carry the momentum anisotropies due to the different 
accelerations of QGP expansion in the transverse plane. This will also be inherited to the final 
charmonium observables~\cite{Chen:2016mhl,Du:2015wha}.

In the semi-central collisions, there are hadronic collisions in the overlap area of two nuclei. 
Meanwhile, nucleus with large electric charges carried by protons moves 
with nearly the speed of light $v_{N}=\sqrt{E_N^2-m_N^2}/E_N > 0.999c$ at RHIC and the LHC energies with the 
relation $E_N= \sqrt{s_{NN}}/2$. The strong electromagnetic (EM) pulse from the moving protons 
can not only change charged/chiral particle evolutions such as 
Chiral Magnetic Effect (CME)~\cite{Kharzeev:2007jp}, etc, but also interact with the target nucleus to generate 
new particles called photoproduction~\cite{Shi:2017qep,Klein:1999qj}. 
This has been widely studied in Ultra-peripheral collisions absent of hadronic 
collisions, and are expected to be negligible with the existence of QGP. However, both recent experimental 
and theoretical 
studies indicate that this photoproduction can be several times larger than the contribution from hadronic collisions, 
but only in the extremely low transverse momentum region $p_T\sim 1/R_A<0.1$ GeV/c~\cite{Adam:2015gba}. 
These photoproduced charmonia from EM fields will also suffer hot medium dissociations generated by the hadronic 
collisions, and in turn gives more constraints on the magnitude of the EM fields and EM-induced particle 
evolutions. 
%\section{Theoretical models}
%\label{sec-theory}

\section{Charmonium from primordial production, regeneration and photoproduction}
\label{sec-PT}
Charmonium production from hadronic collisions and the QGP modifications can 
be well described with microscopic transport models. 
Extensive theoretical and experimental studies indicate multiple mechanisms contributing to the final charmonium 
yield, each of them dominant in different $p_T$ bins see Fig.\ref{fig-schematic}. 
Focus on the primordial production at first, we present charmonium evolutions in p-Pb collisions, where 
regeneration process is suppressed due to the small number of charm pairs in QGP.

\subsection{primordial production and the hot medium effects in pA collisions}
\label{sec-pri}

\begin{figure}
\begin{minipage}[t]{0.33\linewidth}
\centering
\includegraphics[width=1.6in]{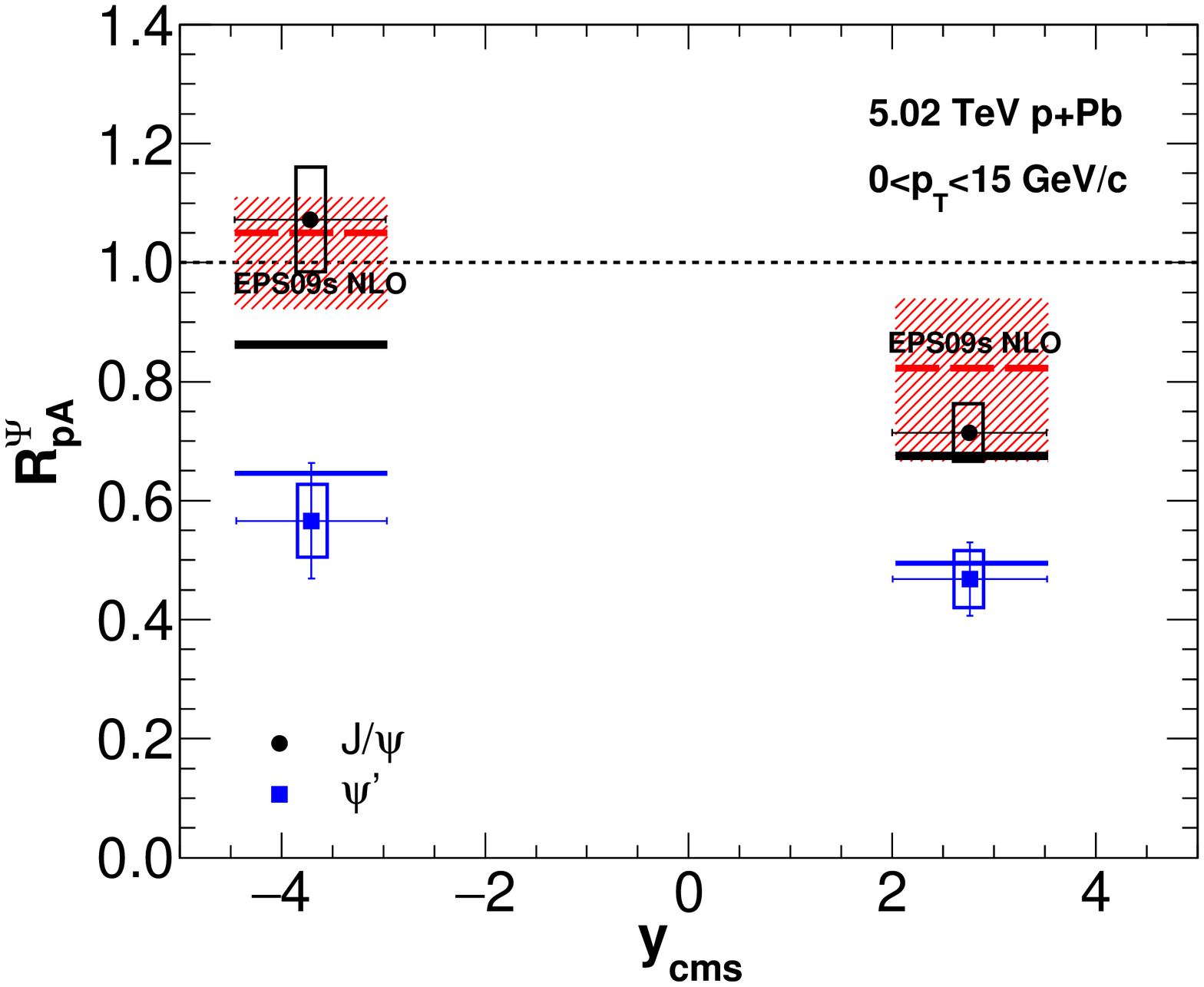}
%\caption{
%Red line represents the situation without shadowing effect. The shadowing effect will reduce charm
%pair by 25\%. }
\label{fig:side:a}
\end{minipage}%
\begin{minipage}[t]{0.33\linewidth}
\centering
\includegraphics[width=1.6in]{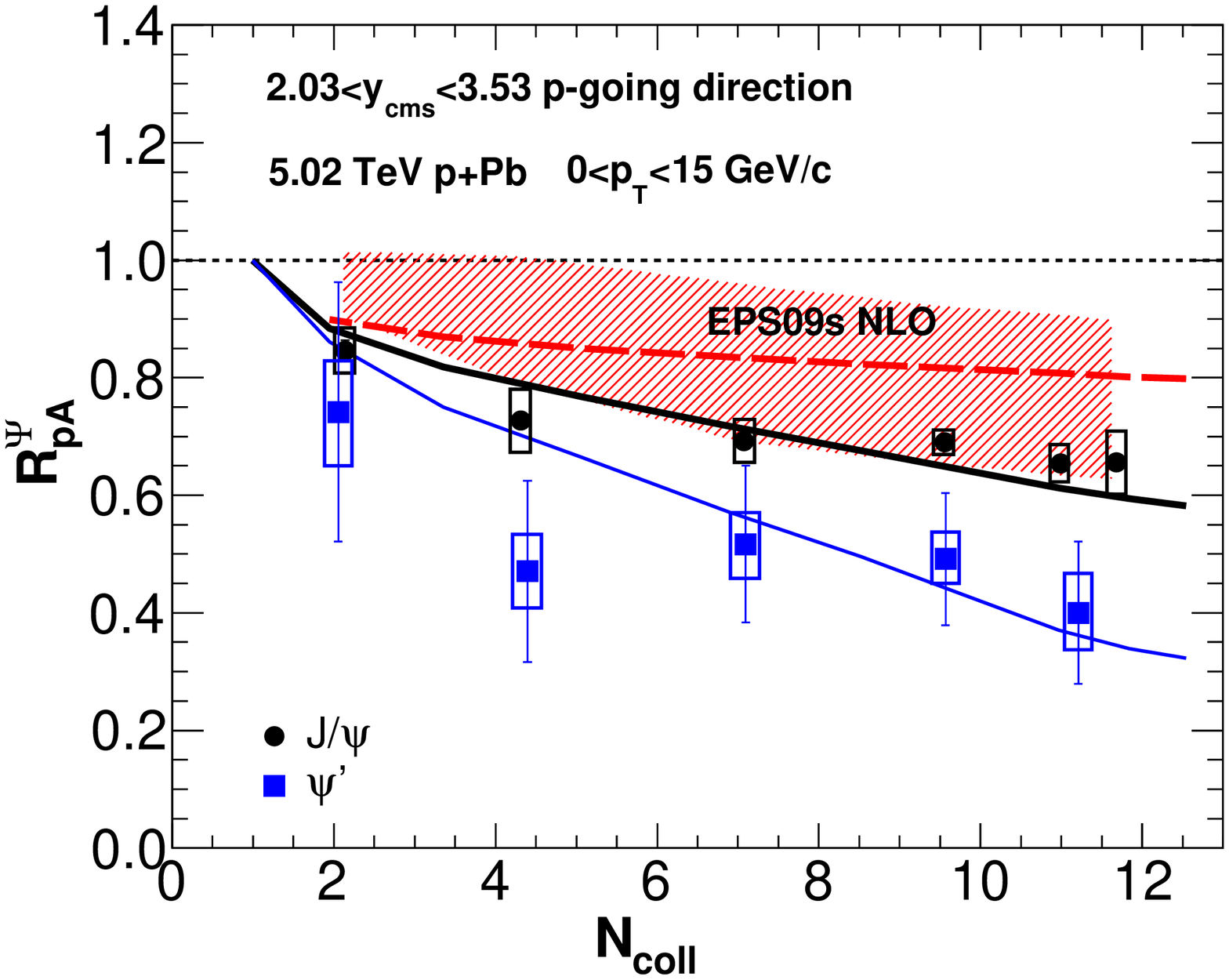}
%\caption{
%forward rapidity. inclusive RAA. }
\label{fig2}
\end{minipage}
\begin{minipage}[t]{0.33\linewidth}
\centering
\includegraphics[width=1.6in, clip]{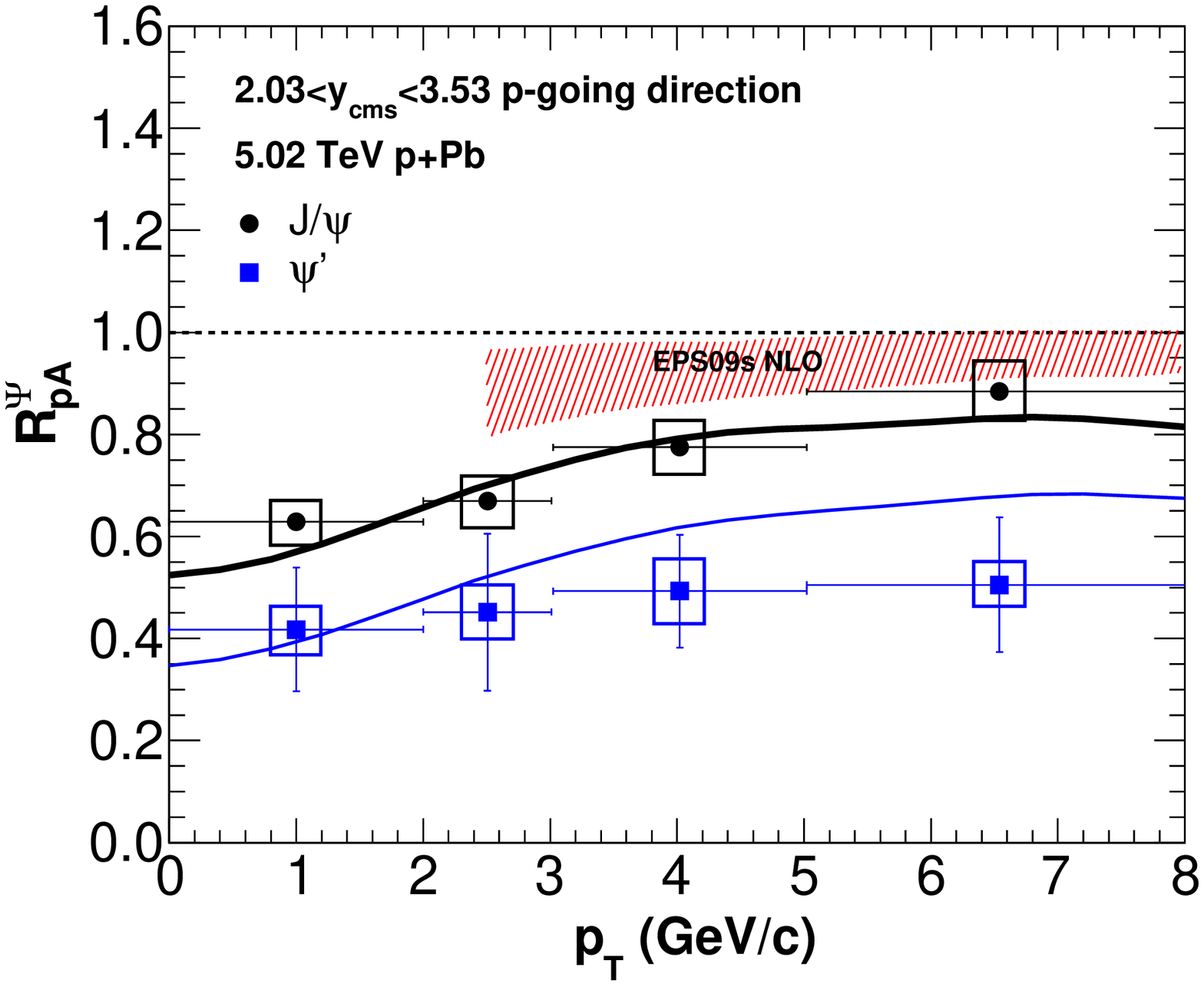}
%\caption{
%forward rapidity. inclusive RAA. }
\end{minipage}
\caption{ (Color online) 
$J/\psi$ and $\psi(2S)$ nuclear modification factors as a function of rapidity, number of binary collisions 
(centralities), transverse momentum respectively at $\sqrt{s_{NN}}=5.02$ TeV p-Pb collisions. The band only includes 
shadowing effects. Lines are the full calculations with both cold and hot medium effects for $J/\psi$ and $\psi(2S)$. 
These figures are cited from~\cite{Chen:2016dke}.   
}
\label{fig-pA}
\end{figure}

\begin{figure}[h]
% Use the relevant command for your figure-insertion program
% to insert the figure file.
\centering
\includegraphics[width=4cm,clip]{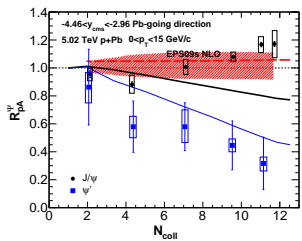}
\caption{$J/\psi$ and $\psi(2S)$ nuclear modification factor as a function of the number of binary collisions in the 
backward rapidity at $\sqrt{s_{NN}}=5.02$ TeV p-Pb collisions. The figure is cited from~\cite{Chen:2016dke}. }
\label{fig-backpA}       % Give a unique label
\end{figure}

We study the nuclear modification factors of different charmonium states which suffer similar cold nuclear matter effects 
(such as shadowing effect on the parton densities in the nucleus) 
but different final state interactions. 
The heavy quark pairs are produced almost instantly 
at $\tau<1/(2m_c)$ where $m_c$ is the mass of charm quark. 
Neglect charmonium formation process and 
assume the heavy quark pair can evolve into a certain 
eigenstate before the time scale $\tau_0\sim 0.6$ fm/c 
of QGP reaching local equilibrium, one can simulate the evolutions of 
charmonium density in phase space in $\tau\ge \tau_0$ with the transport model~\cite{Shi:2017qep}, 
\begin{align}
p^\mu\partial_\mu f_\Psi = - C_\Psi f_\Psi + \beta_\Psi
\end{align}
here $f_\Psi$ is the charmonium phase space density, and $C_\Psi$ is the decay rate caused by hot medium interactions 
including parton inelastic scatterings and color screening. $\beta_\Psi$ will be zero when absent of regeneration. 
Different final state interactions for $J/\psi$ and $\psi(2S)$ are embodied in the factor $C_\Psi$. 
In the hot medium produced in p-Pb collisions, both initial energy density and the geometry size of QGP 
is relatively small compared with Pb-Pb collisions. $J/\psi$ eigenstate can survive from this hot spot, but $\psi(2S)$ 
suffer non-negligible dissociations from the deconfined medium. This makes $R_{AA}^{\psi(2S)}$ becomes smaller than $R_{AA}^{J/\psi}$ 
in different rapidities, centralities and the momentum bin, see Fig.\ref{fig-pA}. The model 
of ``similar cold nuclear matter effects + different QGP suppressions" can explain well $R_{AA}^{J/\psi}$ and 
$R_{AA}^{\psi(2S)}$ in the forward rapidity~\cite{Chen:2016dke}. 
In the backward rapidity defined as Pb-going direction where hotter medium is expected, 
anti-shadowing effect can enhance the $J/\psi$ and $\psi(2S)$ yields at the same time. With only shadowing effect 
from EPS09 NLO model~\cite{Eskola:2009uj}, 
both $J/\psi$ and $\psi(2S)$ $R_{AA}$ is around $\sim 1.1$. However, experimental data 
of $R_{AA}^{\psi(2S)}$ in the 
backward rapidity is only $0.3$ in the central collisions with $N_{coll}\sim 11$. 
After including the hot medium suppressions, 
$R_{AA}^{\psi(2S)}$ from the transport model is close to the experimental data, but $R_{AA}^{J/\psi}$ is also 
suppressed to 0.8 below the experimental data, see Fig.\ref{fig-backpA}. 
This result is consistent with the transport model developed by TAMU Group~\cite{Du:2018wsj}. 
Note that the calculations from Comover model seem to be closer to the experimental data ($\sim 0.95$ in the 
central collisions)~\cite{Ferreiro:2014bia} as they employ a 
larger anti-shadowing effect which can enhance both $J/\psi$ and $\psi(2S)$ yields. We can obtain similar results 
if employing a larger anti-shadowing factor as an input of the transport model. Another possible explanation for 
Fig.\ref{fig-backpA}, is that $\psi(2S)$ transit into $J/\psi$ with in-medium HQ potential 
instead of being dissociated. As at $T\sim T_c$, HQ 
potential from Lattice QCD calculations is restored at the distance of $J/\psi$ mean radius, but is still 
screened at the distance of $\psi(2S)$ radius. This partially screened in-medium HQ potential can result in 
transition of $\psi(2S)\leftrightarrow J/\psi$, and change the relation between $R_{AA}^{J/\psi}$ 
and $R_{AA}^{\psi(2S)}$~\cite{Chen:2016vha}. 

% below is for regeneration, and charm diffusions
\subsection{sequential regeneration and the charm quark diffusion}
\label{sec-rege}

In the Pb-Pb collisions at LHC energies, QGP can dissociate most of the primordially produced charmonia. 
The regeneration becomes dominant in the low $p_T$ region. 
The process of charm quark energy loss and the momentum thermalization in QGP can be studied with Langevin equation, 
transport models, etc. In the limit of kinetic thermalization, charm quark spatial 
densities will be diluted 
by the expansion of the bulk medium~\cite{Zhao:2017yan}. In the early stage of QGP expansion, 
charmonium regeneration is difficult to happen due to the strong color screening on HQ potential. 
Regenerated charmonia can only survive in the later stage of QGP expansions when in-medium heavy quark potential 
is partially restored. In higher colliding energies with larger QGP initial temperature, hot medium takes longer to 
cool down and regenerate charmonia. That means, charm quarks in the expanding QGP will be ``blown" outside and 
their spatial densities is relatively suppressed. Meanwhile, the regenerated charmonia will carry larger collective 
flows and larger momentum~\cite{Chen:2016mhl}. 

\begin{figure}
\centering
\begin{minipage}[t]{0.4\linewidth}
\centering
\includegraphics[height=2.0in]{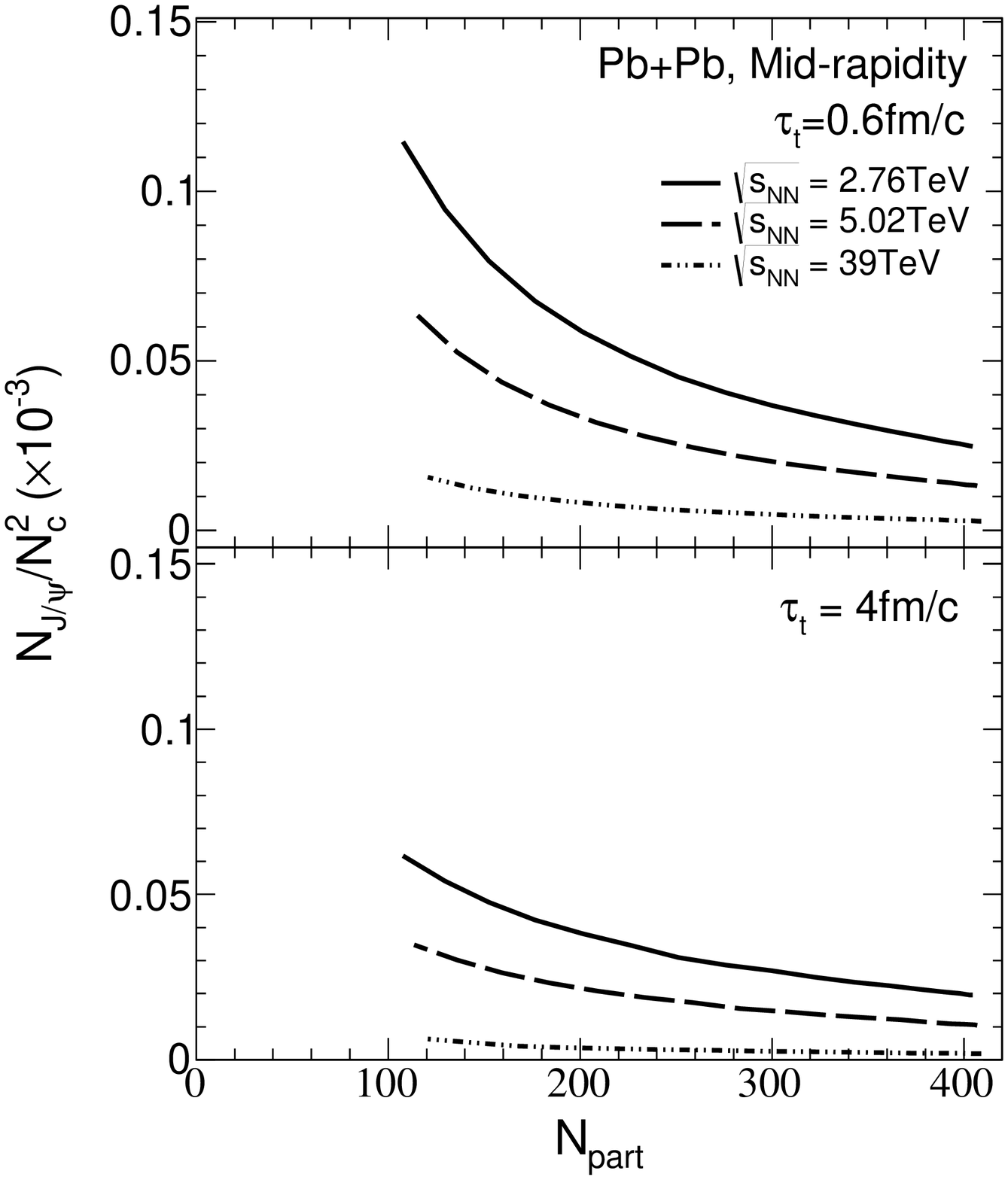}
\end{minipage}%
\begin{minipage}[t]{0.4\linewidth}
\centering
\includegraphics[height=2.0in]{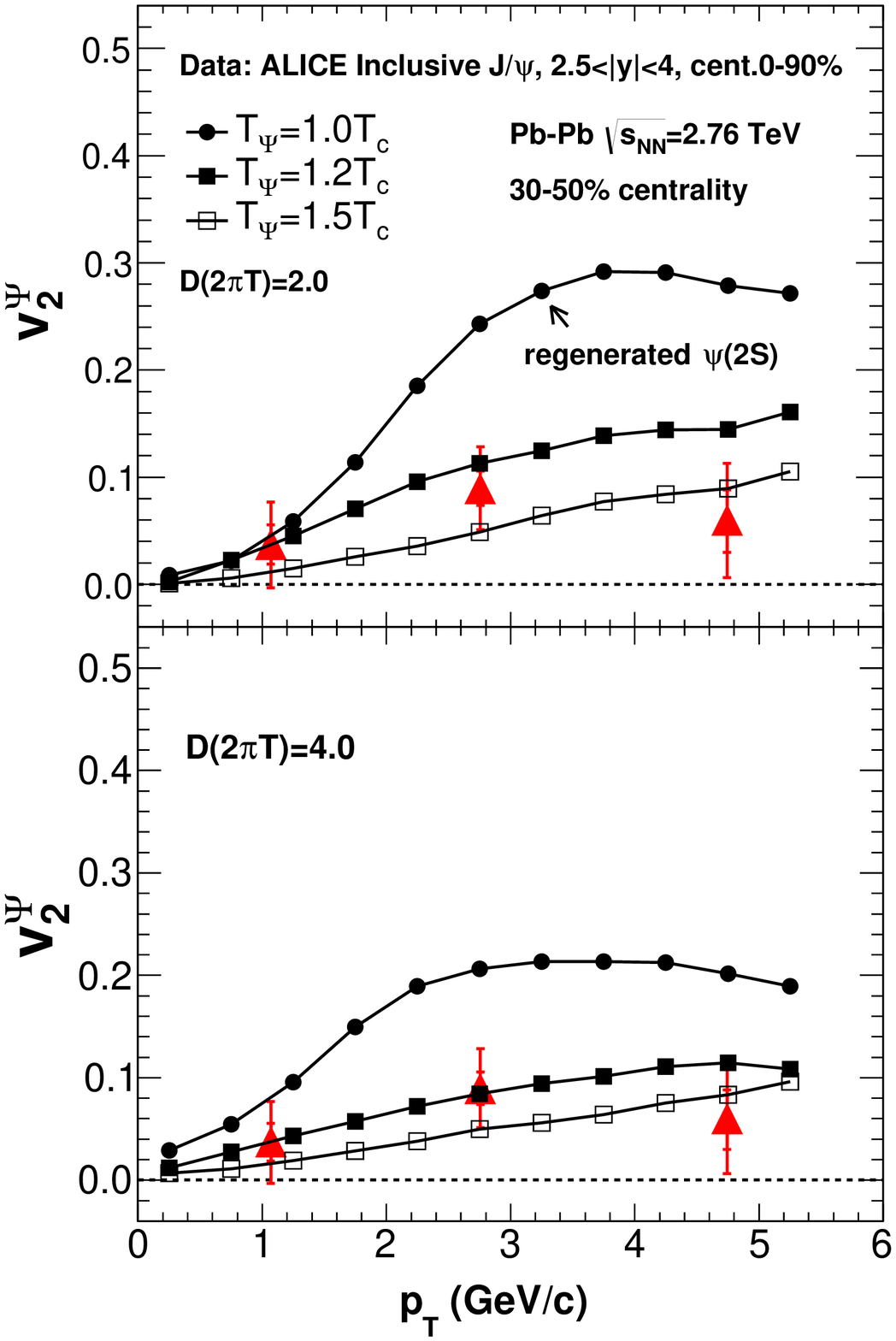}
\end{minipage}
\caption{ The left panel is $N_{J/\psi}/(N_c)^2$ as a function of number of participants $N_p$ with 
the assumption of charm quark 
reaching kinetic equilibrium at different time $\tau_f$, here $N_c$ represents 
equals the number of all charmed particles. The right panel is the elliptic flow of 
regenerated charmonia at different temperature. It is simulated by the Langevin equation.  These figures are 
cited from~\cite{Chen:2016mhl,Zhao:2017yan}. 
}
\label{fig-partonic}
\end{figure}

In order to see the charm quark diffusion effect on regenerated charmonia, 
one can focus on the observable $N_{AA}^{J/\psi}/(N_{c})^2$ (here $N_c$ represents charm quark number in AA collisions). 
It describes the probability of one charm 
combining with another anti-charm quark to generate $J/\psi$. This observable does not depend on 
the shadowing effect and the charm pair cross section $\sigma_{pp}^{c\bar c}$. Its value is mainly determined by 
the evolutions of charm quark density and the recombination rate.  
In the left panel of Fig.\ref{fig-partonic}, from semi-central to central collisions, the initial temperature of 
QGP increases which results in stronger diffusions of charm quarks in the QGP. From 
$\sqrt{s_{NN}}=2.76$ TeV to 39 TeV, the value of 
$N_{AA}^{J/\psi}/(N_{c})^2$ decreases with $\sqrt{s_{NN}}$. 
This charm diffusion effect is also reflected in the comparison of $J/\psi$ and $\psi(2S)$. As $J/\psi$ and 
$\psi(2S)$ are thermally produced in the different stages of QGP expansion due to different binding energies, 
charm quarks can inherit larger collective flows from the anisotropic bulk medium which will also be inherited 
to the regenerated $\psi(2S)$~\cite{Chen:2018kfo}. 
In the right panel of Fig.\ref{fig-partonic}, with the model consisting of  
``Langevin equation for charm evolution + coalescence mechanism for charm hadronization"~\cite{Chen:2016mhl,He:2011qa}, 
elliptic flow of 
regenerated $\psi(2S)$ is always larger than the value of regenerated $J/\psi$. 

\subsection{photoproduction in extremely low $p_T$}

Nucleus moves with nearly the speed of light to generate electromagnetic pulse in the relativistic 
heavy ion collisions. This electromagnetic field interacts with the other nucleus to produce vector mesons or 
dileptons. The photoproduction, 
however, was believed to be negligible compared with the contribution of hadronic collisions. Therefore, the 
photoproduction has been widely studied only in Ultra-peripheral collisions where photoproduction becomes the only 
source for particle production~\cite{Baur:2001jj,Khoze:2002dc,Klein:1999qj,Yu:2017pot,Yu:2017rfi}. 
Now, we find that photoproduction is also visible in the semi-central 
collisions but only in $p_T<0.1$ GeV/c. 

\begin{figure}
\centering
\begin{minipage}[t]{0.33\linewidth}
\centering
\includegraphics[width=1.6in]{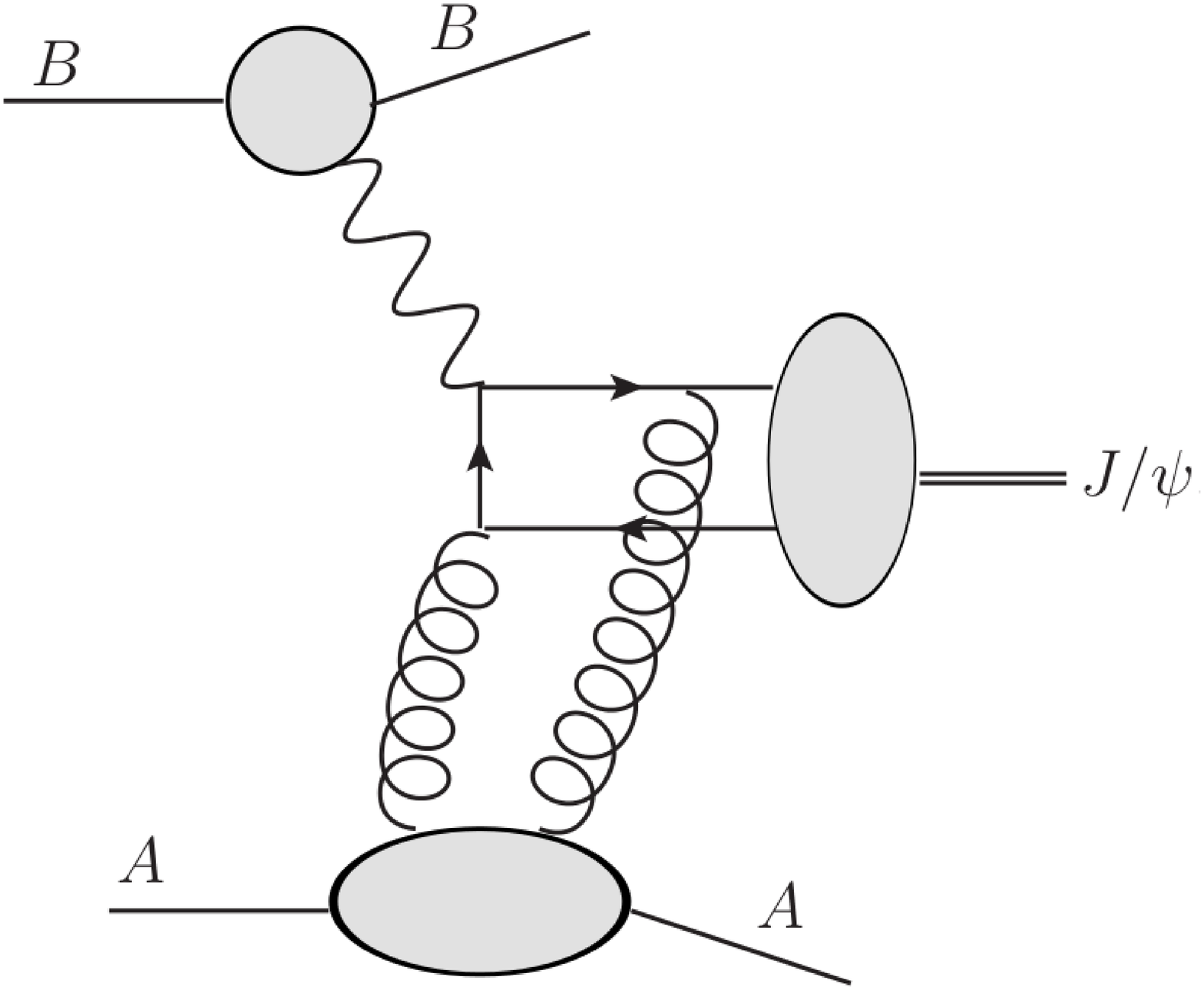}
%\caption{
%Red line represents the situation without shadowing effect. The shadowing effect will reduce charm
%pair by 25\%. }
\end{minipage}%
\begin{minipage}[t]{0.33\linewidth}
\centering
\includegraphics[width=1.6in]{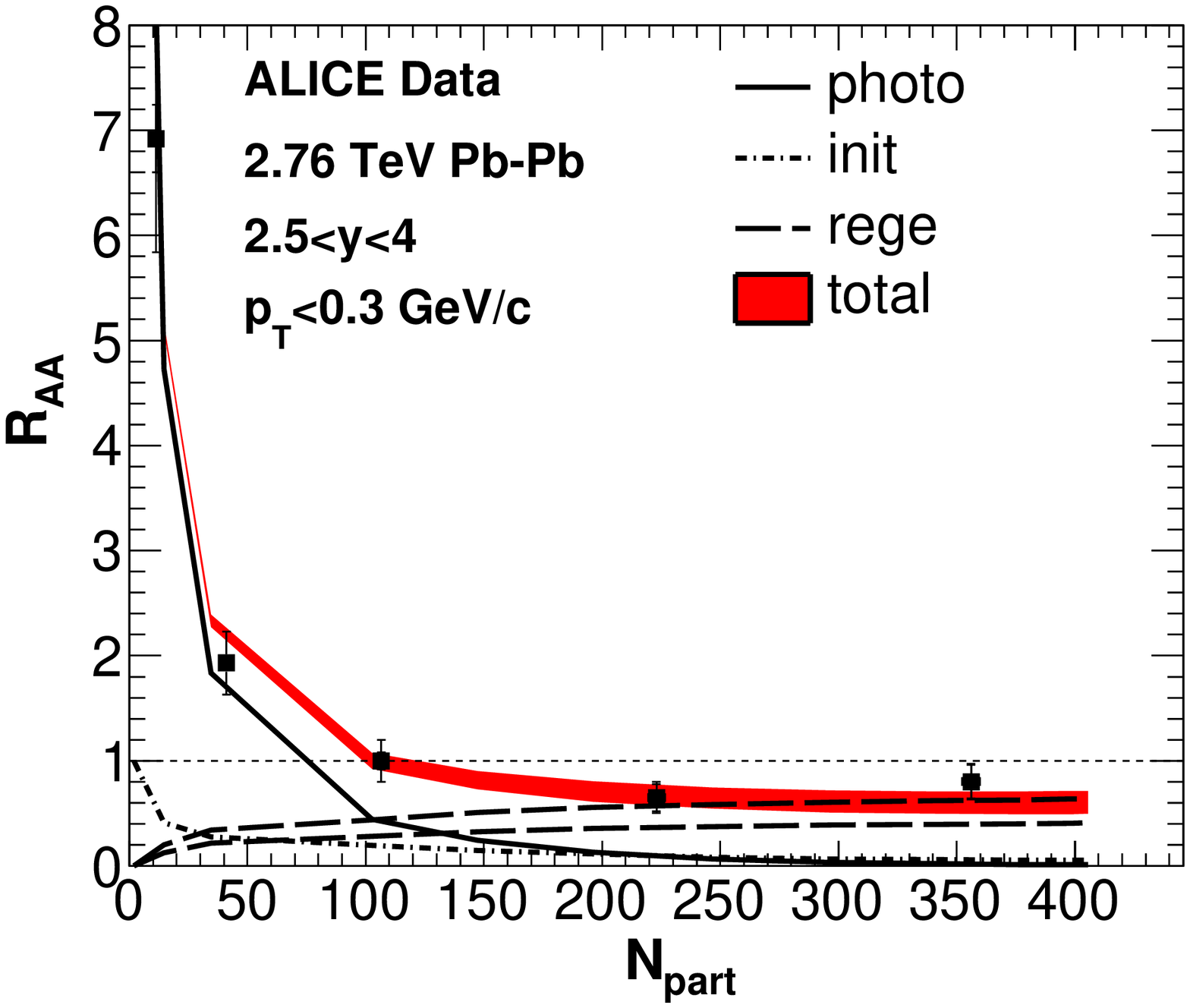}
%\caption{
%forward rapidity. inclusive RAA. }
\label{fig2}
\end{minipage}
\begin{minipage}[t]{0.33\linewidth}
\centering
\includegraphics[width=1.6in]{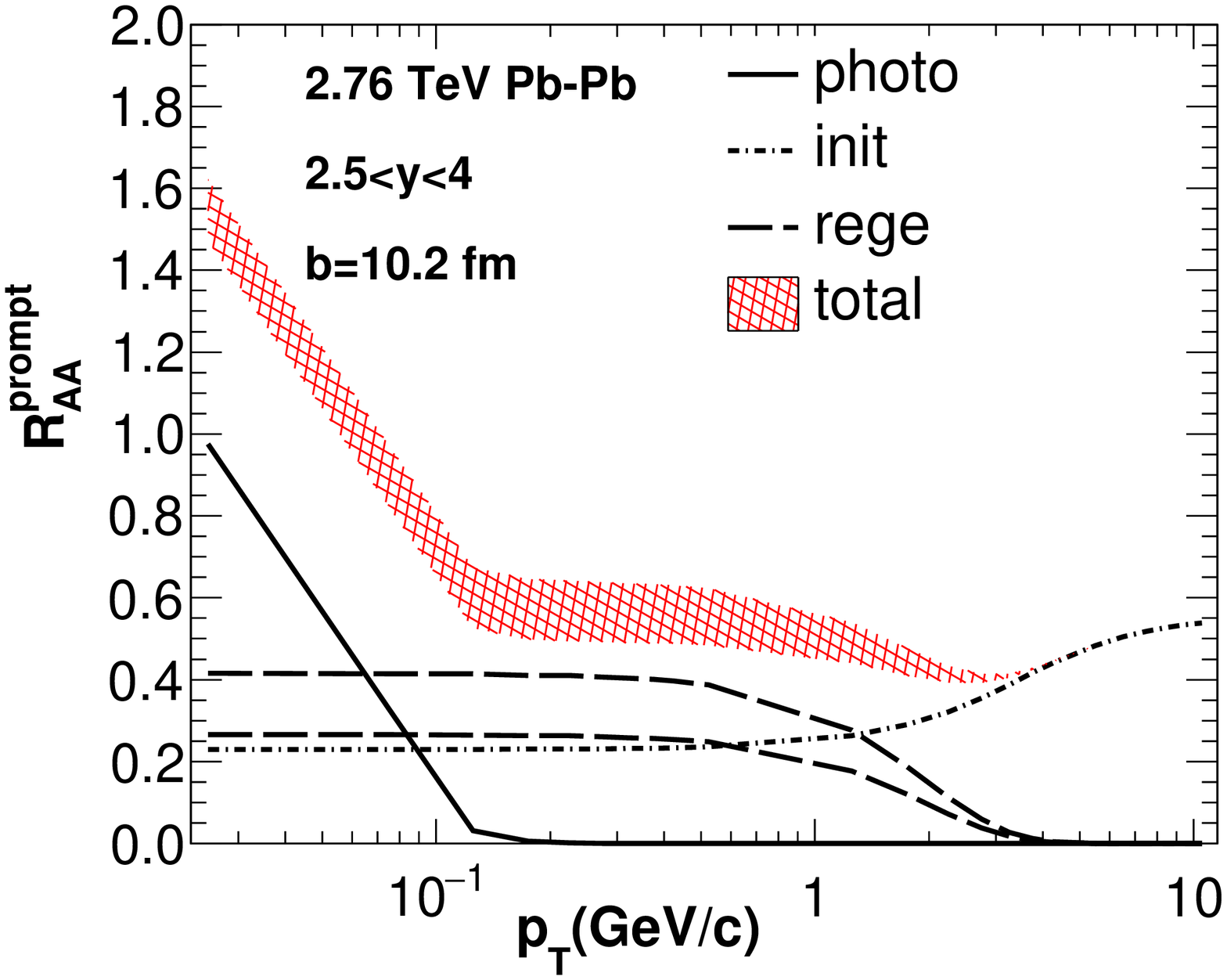}
%\caption{
%forward rapidity. inclusive RAA. }
\end{minipage}
\caption{ (Color online) 
Left panel is the schematic figure for photoproduction. The middle panel is $J/\psi$ inclusive 
$R_{AA}$ as a function of 
$N_p$ in the extremely low $p_T$ at 2.76 TeV Pb-Pb collisions. The right panel is the prompt $R_{AA}$ as a function 
of $p_T$. These figures are cited from~\cite{Chen:2018sir,Shi:2017qep}. 
}
\label{photo-all}
\end{figure}

Instead of dealing with the interactions between electromagnetic fields (generated by one nucleus) 
and the other nucleus, one can approximate this Lorentz-contracted EM fields to be longitudinally moving quasi-real 
photons, see Eq.(\ref{eq-photon})~\cite{Fermi:1924tc}. 
The longitudinal direction is defined as the beam direction. This quasi-real photons can scatter with the 
nucleus to fluctuate into a certain vector meson such as $\phi$, $J/\psi$, $\psi(2S)$, etc, see Eq.(\ref{eq-sigma}) 
and the left panel of Fig.\ref{photo-all}.
As photons interact 
with the target nucleus coherently, the typical transverse momentum of the produced particle is $p_T\sim 1/R_A$, 
\begin{align}
\label{eq-photon}
\int d\tau\int d{\bf x}_T \cdot ({\bf E}_T\times {\bf B}_T) &= \int dw w {dN_\gamma\over dw}  \\
\label{eq-sigma}
{d\sigma_{AA\rightarrow AAJ/\psi}\over dy}(y) &= {dN_\gamma\over dy}(y)\sigma_{\gamma A\rightarrow AJ/\psi}(y) 
+  {dN_\gamma\over dy}(-y)\sigma_{\gamma A\rightarrow AJ/\psi}(-y) 
\end{align}

As the cross section of photoproduction in Eq.(\ref{eq-sigma}) is proportional to the forth of nuclear electric charges 
$\propto Z_e^4$, this photoproduction becomes important in AA collisions but negligible in the pp 
collisions. This additional contribution can enhance the nuclear modification factor $R_{AA}$ in the semi-central and 
peripheral collisions in the extremely low $p_T$ region, see the middle panel of Fig.\ref{photo-all}. Note that in 
the Ultra-peripheral collisions without hadronic collisions, the denominator of $R_{AA}$ scaled from 
pp collisions approaches to zero, but the numerator is non-zero due to the photoproduction, which 
makes $R_{AA} \rightarrow +\infty$ at $N_p\rightarrow 0$. 
In the central collisions with a large number of binary collisions $N_{coll}$, 
hadroproduction increases significantly and photoproduction becomes negligible. 
The feature of photoproduction becomes more clear in the right panel of 
Fig.\ref{photo-all}. In $p_T>5$ GeV/c of finally detected prompt $J/\psi$, 
primordial production becomes the main source. Regeneration becomes 
dominant in the middle and low $p_T$ bins. In the extremely low $p_T<0.1$ GeV/c, 
the photoproduction becomes larger than 
the hadroproduction, see the right panel of Fig.\ref{photo-all}.  

\subsection{B hadron decay}
Final charmonium can come from the direct production 
and decay from higher excited state, and also the decay of B hadrons. 
This decay contribution from bottom flavor is labelled as ``non-prompt" charmonium. The contributions of primordial 
production and regeneration is classified as the prompt part. 
In pp collisions, the fraction 
of non-prompt part in inclusive yield have been measured by experiments, fitted to be 
$f_B(p_T)=0.04+0.023p_T/(GeV/c)$ for $J/\psi$ and $f_B(p_T) = 0.11+0.022p_T/(GeV/c)$ for $\psi(2S)$. 
The momentum dependence of the $f_B(p_T)$ shows weak dependence 
on the colliding energies and the rapidities~\cite{Aaltonen:2009dm}. 
In AA collisions, the momentum distribution of bottom quarks will be modified 
by QGP when they move through the medium. 
The momentum shift of bottom quarks 
before hadronization changes the non-prompt yield at different $p_T$ regions. 
This hot medium modification factor is called quenched factor $Q(p_T)$.  

\begin{figure}
\centering
\begin{minipage}[t]{0.33\linewidth}
\centering
\includegraphics[height=1.3in]{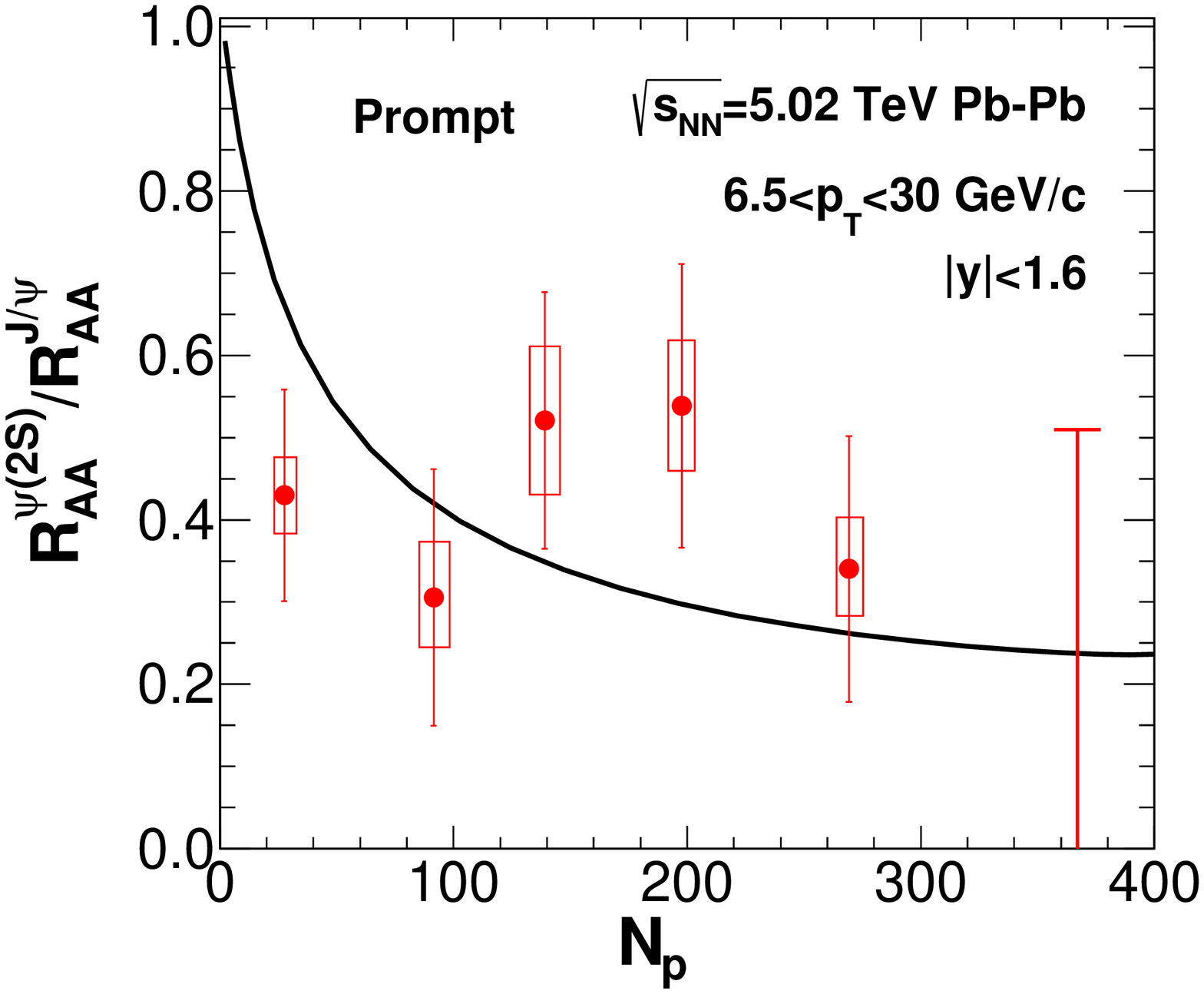}
%\caption{
%Red line represents the situation without shadowing effect. The shadowing effect will reduce charm
%pair by 25\%. }
\label{fig:side:a}
\end{minipage}%
\begin{minipage}[t]{0.33\linewidth}
\centering
\includegraphics[height=1.3in]{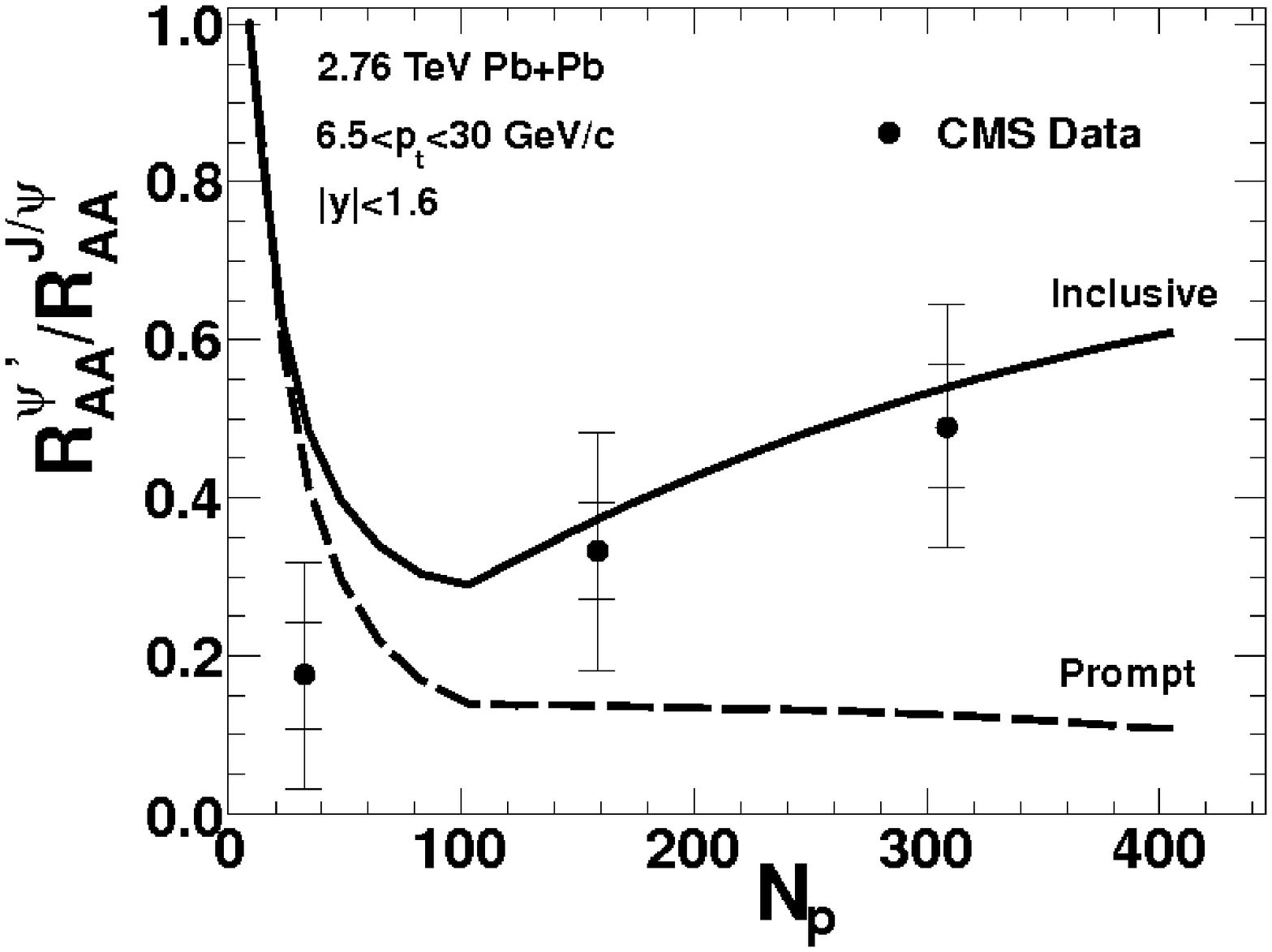}
%\caption{
%forward rapidity. inclusive RAA. }
\label{fig2}
\end{minipage}
\begin{minipage}[t]{0.33\linewidth}
\centering
\includegraphics[height=1.3in]{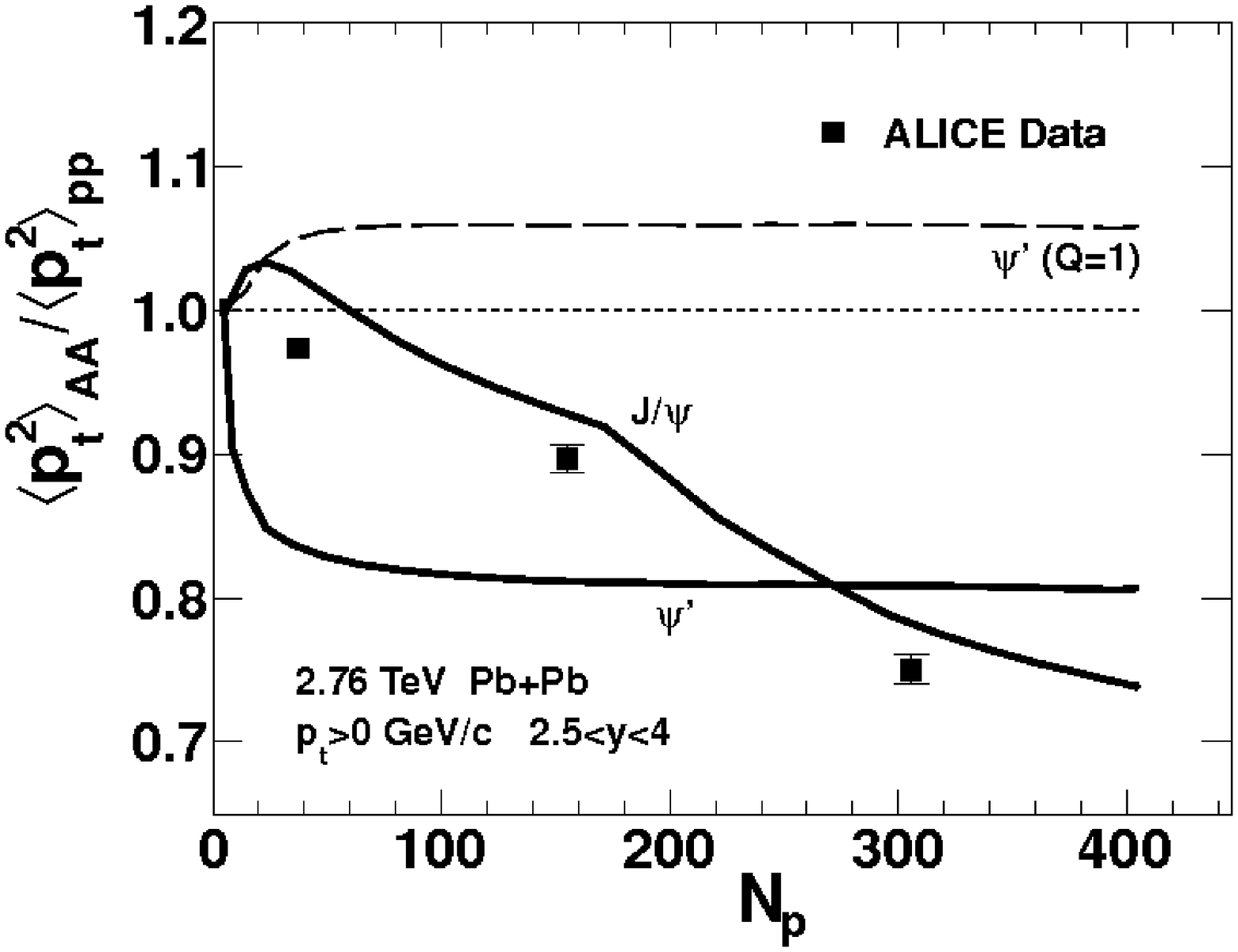}
%\caption{
%forward rapidity. inclusive RAA. }
\label{fig2}
\end{minipage}
\caption{ (Color online) 
The left panel is prompt $R_{AA}^{\psi(2S)}/R_{AA}^{J/\psi}$ in $6.5<p_T<30$ GeV/c in $\sqrt{s_{NN}}=5.02$ TeV Pb-Pb 
collisions. The middle panel is prompt (dashed line) and inclusive (solid line) $R_{AA}^{\psi(2S)}/R_{AA}^{J/\psi}$ 
at $\sqrt{s_{NN}}=2.76$ TeV. The decay rates for charmonia in left panel is updated compared with the middle panel. 
The right panel is $\langle p_T^2\rangle_{AA}/\langle p_T^2\rangle_{pp}$ of inclusive $J/\psi$ and 
inclusive $\psi(2S)$. These figures are cited from AAAA.  
}
\label{fig-Bdecay}
\end{figure}

When we focus on prompt $R_{AA}$ of $J/\psi$ and $\psi(2S)$ at $p_T>6.5$ GeV/c, 
lack of regeneration in this high $p_T$,  
primordially produced $\psi(2S)$s are more easily dissociated by QGP compared with 
$J/\psi$, 
$R_{AA}^{\psi(2S)}/R_{AA}^{J/\psi}$ decreases with $N_p$, see the left panel of Fig.\ref{fig-Bdecay}. 

After including the contribution of B hadron decay, $J/\psi$ yield is less affected, as non-prompt $J/\psi$ 
is less than the prompt ones due to $J/\psi$ large binding energy. 
However, prompt $\psi(2S)$s in the primordial 
production are strongly dissociated, their final inclusive yield is then dominated by the non-prompt part, see 
the second term of Eq.(\ref{eq-RAA}). 
$r_B= f_B/(1-f_B)$ represents the ratio of non-prompt and prompt charmonium yields. 
The quench factor $Q$ shows flat tendency with $N_p$ in the region $100<N_p<400$ in experiments~\cite{Chen:2018sir}. 
This makes inclusive $R_{AA}^{\psi(2S)}$ become flat.  Inclusive $R_{AA}^{J/\psi}$ is dominated by 
the first term of Eq.(\ref{eq-RAA}) and decrease with 
$N_p$. Therefore, the inclusive 
$R_{AA}^{\psi(2S)}/R_{AA}^{J/\psi}$ decrease at first and then show clear increasing tendency at $100<N_p<400$, 
see the middle panel of Fig.\ref{fig-Bdecay}. 
\begin{align}
R_{AA}^{\rm inclusive}= R_{AA}^{\rm prompt}{1\over 1 + r_B} + {r_B\over 1+r_B} Q
\label{eq-RAA}
\end{align} 
In the entire $p_T$ region, regenerated $J/\psi$ with small momentum 
will dominate the inclusive production, which suppress $\langle p_T^2\rangle_{J/\psi}$. 
For inclusive $\psi(2S)$, its yield 
mainly consists of the non-prompt part from B hadron decay, as 
prompt part can barely survive from the QGP. This makes 
$\langle p_T^2\rangle_{\psi(2S)}$ become flat with $N_p$, and is 
closely connected with the energy loss of bottom quarks, see the right panel 
of Fig.\ref{fig-Bdecay}.

\section{Summary}
\label{sec-sum}
In summary, we discuss different production mechanisms for both $J/\psi$ and $\psi(2S)$ including 
primordial production, recombination of 
uncorrelated charm quarks in QGP and photoproduction from strong electromagnetic fields. Each of the production process 
dominates charmonium final yield in different $p_T$ region because of their dynamical process. 
Transport models with all above ingredients can capture the main features of the experimental data of $J/\psi$ 
and $\psi(2S)$ in both p-Pb and Pb-Pb collisions. These studies can provide clear picture of heavy quarkonium 
evolutions in the cooling deconfined medium. With the significant contribution of regeneration, hidden and open charm 
mesons are connected to each other. The space-time dependence of heavy quark diffusions also leave 
imprints on the heavy quarkonium yields and momentum distributions. The precise measurements about charmonium production 
in extremely low $p_T$ also connects the hadroproduction from hadronic collisions with the photoproduction from 
electromagnetic fields which acts as one of the most important inputs in the studies of chiral/charged particle 
evolutions in heavy ion collisions. More explicit experimental data in small and 
large colliding systems for different charmonium states 
also provide promising opportunities to study the 
quantum effects of quarkonium states in the thermal medium. 

\vspace{0.5cm}
{\bf Acknowledgements} 
We thank Prof. Pengfei Zhuang for the help in above studies. 
B. Chen has been supported by
NSFC Grant No. 11705125 and Sino-Germany (CSC-DAAD)
Postdoc Scholarship.

\end{document}